\documentclass[graybox]{svmult}

\usepackage{type1cm}        
\usepackage{makeidx}         
\usepackage{graphicx}        
                             
\usepackage{caption}
\usepackage{subcaption}
\usepackage{multicol}        
\usepackage[bottom]{footmisc}

\usepackage{newtxtext}       %
\usepackage{newtxmath}       

\makeindex             

\begin{document}

\title*{On Connectivity-Aware Distributed Mobility Models for Area Coverage in Drone Networks}
\author{Mustafa Tosun \and
Umut Can Çabuk \and
Vahid Akram \and
Orhan Dagdeviren}
\institute{Mustafa Tosun \at Ege University, Izmir, Turkey, \email{mustafa.tosun@ege.edu.tr}
\and Umut Can Cabuk \at Ege University, Izmir, Turkey, \email{umut.can.cabuk@ege.edu.tr}
\and Vahid Akram \at Ege University, Izmir, Turkey, \email{vahid.akram@ege.edu.tr}
\and Orhan Dagdeviren \at Ege University, Izmir, Turkey, \email{orhan.dagdeviren@ege.edu.tr}
}
%
%
\maketitle

\abstract{Drone networks are becoming increasingly popular in recent years and they are being used in many applications such as area coverage, delivery systems, military operations, etc. Area coverage is a broad family of applications where a group of connected drones collaboratively visit the whole or parts of an area to fulfill a specific objective and is widely being researched. Accordingly, different mobility models have been designed to define the rules of movements of the participating drones. However, most of them do not consider the network connectivity which is crucial, plus many models lack the priorities and optimization strategies that are important for drone networks. Therefore within this study, three known connectivity-aware mobility models have been analyzed comparatively. Two non-connectivity-aware mobility models have further been implemented to catch the placebo effect if any. Per the detailed experiments on the mobility models, coverage rates, connectivity levels, and message traffic have been evaluated. The study shows that the Distributed Pheromone Repel (DPR) model provides a decent coverage performance, while the Connectivity-based model and the Connected Coverage model provide better connectivity and communication quality.
}

\keywords Area Coverage, Distributed Algorithms, Mobility Models, Connectivity, Drone Networks, UAV.

\section{Introduction}
Recent advances on the communication technologies and the evolution of small size, low cost and low energy hardware have triggered the boost of a new generation of devices that are going to make a revelation on human life. Unmanned aerial vehicles (UAV), especially the multi-rotor aircrafts that are called drones, are one of the future technologies which are increasingly being used in commercial, military, healthcare, transportation, monitoring, security, entertainment, and surveillance applications. UAV’s can also fly autonomously and/or in groups, called swarms, to perform such missions \cite{7463007}. 

Due to the wide range of popular applications and particular challenges, many recent research have focused on UAVs. One obvious drawback of using UAVs is their production/sales costs. Generally, UAVs with wide communication ranges, long flight times, powerful processors and different sensor/actuator equipment are very expensive. The cost constraints, limits the usability of such UAVs, usually to the military applications and make them a less profitable approach for many commercial applications. Using a large swarm of low-cost UAVs, instead of a single (or a few) expensive UAV, is a worthwhile and feasible alternative that has many advantages, such as agility, cost-efficiency and reliability over using a single UAV. By utilizing a swarm, the joint coverage area can easily be extended via adding more UAVs to the network, the cooperative task can be done by collaboration of UAVs and persistent services can be guaranteed by periodically replacing the worn-out UAVs. However, the swarms, have also some challenges, such as communication issues, limited range, mobility control, path planning and dynamic routing. 

The maximum radio range of inexpensive UAVs are typically limited, which eventually restricts the allowable distances to the neighboring UAVs. Therefore, each flying UAV should remain in the range of at least one other UAV, which is connected to some other UAVs. In case of disconnections due to limited ranges, it is proved that the restoring the connectivity between the UAVs in a disconnected drone network is an NP-hard problem \cite{akram2018hardness}, hence to avoid losing the resources the drones should always be kept connected to the rest of the swarm. To face this problem, many research have proposed different mobility models that describe how the drones should change their position while they communicate with other drones in their radio range. The mobility of flying drones in 3 dimensional space is completely different from terrestrial vehicles, which mostly utilizes the 2 dimensional space. Hence, the available mobility models for traditional mobile ad-hoc networks such as the Random Waypoint (RWP) model \cite{hyytia2005random} are not neatly applicable to UAV swarms. In the case of swarms, mobility models that are specifically designed for flying objects are required. 

In this paper, the selected mobility models for flying ad-hoc networks (FANET) have been described and further analyzed to evaluate their performances from various perspectives. Five of the notable mobility models for FANETs were implemented on a discrete event simulator and their performances were compared accordingly, using different metrics including the speed of coverage (convergance rate), the fairness of coverage, the connectivity percentage (convergance degree), the average connected components throughout the mission and the required message traffic volume. Briefly, the main contributions of this work are as follows:
\begin{itemize}
    \item A review highlighting the advantages and the disadvantages of available mobility models for FANETs consisting of rotating-wing drones has been provided.
    \item Five notable mobility models have been implemented successfully in Python using the SimPy library. 
    \item The performance of implemented models have been comparatively evaluated using five different metrics (listed above). 
\end{itemize}

The rest of the paper has been structured as follows: Section II presents a review about the available mobility models for FANETs. Section III describes the details of the five selected mobility models. Section IV explains the important metrics that can be used to analyze the performance of mobility models. Section V provides the simulation results of the implemented mobility models and compares the performances, thereof. Finally, Section VI draws a conclusion and points out the potential future works.  

\section{Related Works}
Various mobility models for UAV swarms have been proposed by different studies, in which each model has focused on specific properties such as application scenarios, area coverage, connectivity, speed and/or particular UAV types. Most papers consider the fixed-wing UAVs, while rotating-wing UAVS (drones) are also researched in some recent works.

The self-organizing aerial ad-hoc network mobility model proposed by Sanchez et al. in \cite{sanchez2015self} focused on disaster recovery applications. In this model, the UAVs keep their connectivity alive by pushing and pulling each other to some extent according to their similarities and a threshold value. This model focuses on victim spotting and saving scenarios, but ignores the area coverage constraints.

Kuiper and Tehrani has proposed a mobility model, named pheromone, for UAV groups in reconnaissance applications \cite{kuiper2006mobility}. This model only considers the coverage constraint and ignores the connectivity issues. More, the model has only three directions which is only suitable for fixed-wing UAVs. 
To preserve the stability of the connections in autonomous multi-level UAV swarms performing wide-area applications, a connectivity maintenance strategy, which combines the pheromone-based mobility model and some mobile clustering algorithms has been proposed by Danoy et al. in \cite{danoy2015connectivity}. Yanmaz has further investigated the coverage and connectivity of autonomous UAVs in monitoring applications and proposed a connectivity-based mobility model to sustain the connectivity between the UAVs and also to the base station \cite{yanmaz2012connectivity}. The proposed model focuses on connectivity and tries to maximize the covered area by a swarm of connected drones.  

A distributed mobility model for autonomous interconnected UAVs for area exploration has been proposed by Messous et al. in \cite{messous2016network}. In this model, the UAVs, explore a target area while they maintain the connectivity with their neighboring UAVs and the base station. This model takes the energy consumption constraint into account to increase the overall network lifetime, which is a noteworthy feature. Another decentralized and localized algorithm for mobility control of UAVs has been proposed by Schleich et al. in \cite{schleich2013uav}. The proposed algorithm is designed to perform surveillance missions while preserving the connectivity. The UAVs maintain the connectivity between themselves by establishing a tree-based overlay network, where the root of the tree is the base station. However, fully ad-hoc networks that do not include a base station would be out of the scope. A survey about the available mobility models for different applications regarding flying ad-hoc networks has been presented by Bujari et al. in \cite{bujari2017fanet}. This survey compares the models only from the application perspective and does not consider the other (i.e. performance) metrics. 

\section{Mobility Models}
Within the scope of this study, the following models have been implemented: the random-walk as in \cite{kuiper2006mobility}), the Distributed Pheromone Repel model \cite{kuiper2006mobility}, the Connectivity-based model \cite{yanmaz2012connectivity}, the KHOPCA model \cite{danoy2015connectivity}, and the Connected Coverage model \cite{schleich2013uav}. This section briefly explains the details of each model.   

\subsection{Random Mobility Model}
The random mobility (also called random-walk) is a very simple approach, in which the UAVs move independently and randomly in the target area. While there are already more than a few random-walk models in the literature, the variations can even be increased by extending them with suitable connectivity maintenance algorithms  \cite{ijamec264942}. In the considered basic model, UAVs fly in the area in a randomly chosen direction for each second (or for another pre-defined time slot). To change the direction in the target area, the authors in \cite{kuiper2006mobility} have proposed three actions being "go straight", "turn left" and "turn right". UAVs can decide on their actions randomly, based on fixed probabilities. The same strategy has been implemented in this study. The random mobility model is commonly used in many applications such as search and explore missions and monitoring or surveillance operations. The random model is also widely used as a benchmark indicator in many studies.

\subsection{Distributed Pheromone Repel Mobility Model}
In this model the drones may take the same three actions as in the random-walk model which are "go straight", "turn left" and "turn right" \cite{kuiper2006mobility}. However, instead of fixed probabilities for these decisions, in the pheromone-based model, a so called "pheromone" map is used to guide the UAVs. The pheromone map contains information regarding the visited/explored areas and is shared/synchronised among the UAVs. An aircraft exchanges information about the scanned area, and according to this information, the other UAVs decide to turn left, right or go straight ahead.

\subsection{Connectivity-based Mobility Model}

The connectivity-based mobility model is a self-organizing model, in which the UAVs use local neighborhood information to sustain the connectivity of the network \cite{yanmaz2012connectivity}. The UAVs can enter or leave the network at any time. The target area can also be changed during the algorithm execution. The main goal is to cover the maximum possible area, while preserving the connectivity between all UAVs and the base station. Each UAV explores its neighbors and computes its location periodically. Then it decides if it needs to change its direction. The information that is exchanged between the UAVs contain the current locations and the directions.

In every $t$ seconds, each UAV checks whether it is located within the transmission range of the sink UAV. If it has a live communication link to the sink, it checks if it still would be in the range after $t$ seconds, given its current direction and the speed. If the UAV estimates that it will be in the range of the sink, it keeps its previous flying direction. Otherwise, the UAV changes its direction randomly towards a point in the transmission range of the sink. If the sink is not in the radio range, then the UAV tries to find a multi-hop path to the sink. If it can find such a path, it estimates the next location of itself and its neighbors and determines whether it would still be connected to the sink after $t$ seconds. If the path still exists, it keeps its previous direction. If the UAV estimates that the connection will not exist in the next $t$ seconds, then it changes its direction such that it stays connected to a neighbor UAV that provides a multi-hop path to the sink. If a UAV can not find such a path to the sink, it tries to keep the connection to a neighboring UAV to avoid a complete isolation. Finally, when a UAV becomes isolated, it keeps its previous direction until reaching the boundary of the mission field or until meeting another UAV. At the boundaries of the field, the UAV changes its direction randomly towards a point in he field.

\subsection{KHOPCA-based Mobility Model}
This model combines the ACO-based mobility model given in \cite{kuiper2006mobility} with the KHOPCA clustering algorithm presented by Brust et al. in \cite{brust2007adaptive} for the UAV swarms to optimize their surveillance capabilities, coverage area and stability of the connectivity. Each UAV locally executes the KHOPCA-based clustering algorithm. The UAVs select their directions using the pheromone-based approach and the KHOPCA-based probability. The cluster-head UAVs or the UAVs with a lower weight than the KHOPCA threshold  value, uses the pheromone-based mobility model to select their direction. For the other UAVs, the location of a neighbor with the lowest KHOPCA weight is selected as the destination. 

\subsection{Connected Coverage Mobility Model}
This model has three sequential steps as follows: 1- Each UAV selects a preferred neighbor UAV among its one-hop neighbors to stay connected with them. 2- The UAVs compute their current and future positions using the direction and speed values, alternative directions are also computed to stay connected with one of the selected neighbors. 3- A pheromone-based behaviour helps selecting the best direction from the previously computed set. This model maintains a tree structure, where the root is the base station. To compute the alternative directions, each UAV estimates the upcoming position of each neighbor UAV using their current positions and the current destinations of every single neighbor. Yet, the UAVs check where they would be in this predefined future and decide a correct direction to satisfy the connectivity constraint of the model. In this model, UAVs regularly drop virtual pheromones on their ways and these pheromones gradually disappear after some time. Thus, the UAVs will be "attracted" to the places that are marked by the lowest pheromone concentration.

\section{Performance Metrics}

This section introduces the performance metrics that show the quality or usefulness of the models considering the area coverage features, connectivity maintenance and communication costs. The speed of coverage (convergence rate) and the fairness of coverage have been measured to see the quality of the area coverage capabilities. Also, the ratio of the connected UAVs, the ratio of connectivity to the root and the average connected component number shows how well connected the topologies are, by using each mobility model. Finally, one of the most important metrics, namely the communication costs, is calculated. The total number of the sent messages (by each UAV) and the size of these messages are taken into consideration in that calculation.

\emph{Speed of Coverage}: Also the convergence speed or rate. In most area coverage missions, mobility models need to cover the whole area as fast as possible. This metric shows how much time required to scan the specific percentage of the mission field. The coverage time is getting higher and higher as the desired percentage (threshold) increases. Hence, instead of the whole area, the time passed until covering the 80\% and 95\% of the field is recorded to find out the coverage speeds.

\emph{Fairness of Coverage}: The fairness of coverage is the metric that shows how balanced the whole area is scanned. To compute the fairness, it is necessary to record for how many seconds (or times) each square meter (or unit area in general) is already scanned by the UAVs. If two or more UAVs scan the same part of the field (i.e. a cell) at the same second, it should be counted as one second. Hence, the coefficient of variations gives how balanced the scanning times are. It's expected that the fairness to be close to 0.

\emph{Connected Topology Percentage}: Connected topology percentage is the ratio of the non-continuous cumulative time period, in which all the UAVs are connected to the network, to the total time period, which the mobility model needs to cover the 95\% of the mission field. To compute that percentage, the number of the connected components are gathered at each second of the execution time, then the ratio to the execution time is calculated.

\emph{Connectivity Percentage To Root}: This metric is taken into account in the Connectivity Based and the Connected coverage model. Since the drone network applications use the one UAV as a gateway and main processor, these models tries to make drones are connected to the root during execution. Its computed as the average of the connectivity percents of the drones to the root (single-hop or multi-hop).

\emph{Average Connected Component}: This metric is the mean of the connected component numbers which gathered at each second. It shows how much the topology is dispersed during the execution time.

\emph{Message Count}: Message count is the number of messages sent by UAVs during the execution of the mission. This metric was not considered in the related previous works. It gives an insight regarding the communication costs of the models. As the message counts increase, the energy required to send the messages increases, too. It's expected to be lower as a quality/performance indicator.

\emph{Total Message Size}: The message counts are not merely enough to find out the whole communication costs. Because, the energy spent by UAVs (during the communications) is also related to the size of the sent (and received) messages. The message size is directly proportional to the communication cost. Thus, the message counts should be multiplied by the average message sizes for each model throughout the mission to get comparable values.  

\section{Simulation Results}

The mobility models are implemented using Python and the SimPy library \cite{simpy}. The parameters of the simulations are given in Table \ref{tbl1}. The size of the area is defined as 2000 m X 1000 m in a rectangular shape. For the pheromone-based models, the map's cell size is set as 5 m. However, the scan times have been measured for each square meter to get more precise results. The UAVs' radio ranges and operational (e.g. sensor) ranges are fixed to 400 m and 20 m respectively. The programmed autopilot mimics a simple rotating-wing drone behaviors. Accordingly, the speeds of the UAVs are fixed at 5 m/s. The decision intervals are specified differently for each model as specified in their original papers. The communication infrastructure is assumed to be optimal (no loss, no collision, unlimited bandwidth etc.) as in \cite{kuiper2006mobility,schleich2013uav,danoy2015connectivity}. The number of UAVs used varies between 4 to 50 to test the models in sparse and dense environments. Each model was simulated 30 times to get statistically consistent results.

\begin{table}
\caption{Simulation Parameters}
\begin{center}
\begin{tabular}{ l l }
\hline
Parameter name & Parameter Value \\
\hline
Simulation Area \\
Size & 2000 m X 1000 m \\
Map's cell size & 5 m X 5 m \\
Measurement cell size & 1 m X 1 m \\
UAV \\
Speed & 5 m/s \\
Radio range & 400 m \\
Coverage range & 20 m \\
Decision intervals \\
Random & 1 s \\
DPR & 10 s \\
Conn. Based & 2 s \\
KHOPCA Based & 30 s and 0.2 probability \\
Connected Coverage & 30 s \\
Experiments \\
\# of UAVs & [4,6,8,10,15,20,30,40,50] \\
\# of runs per experiment & 30 \\
 \hline
\end{tabular}
\label{tbl1}
\end{center}
\end{table}

\begin{figure}[h!]
    \centering
    \begin{subfigure}{0.5\textwidth}
        \centering
        \includegraphics[scale=0.32]{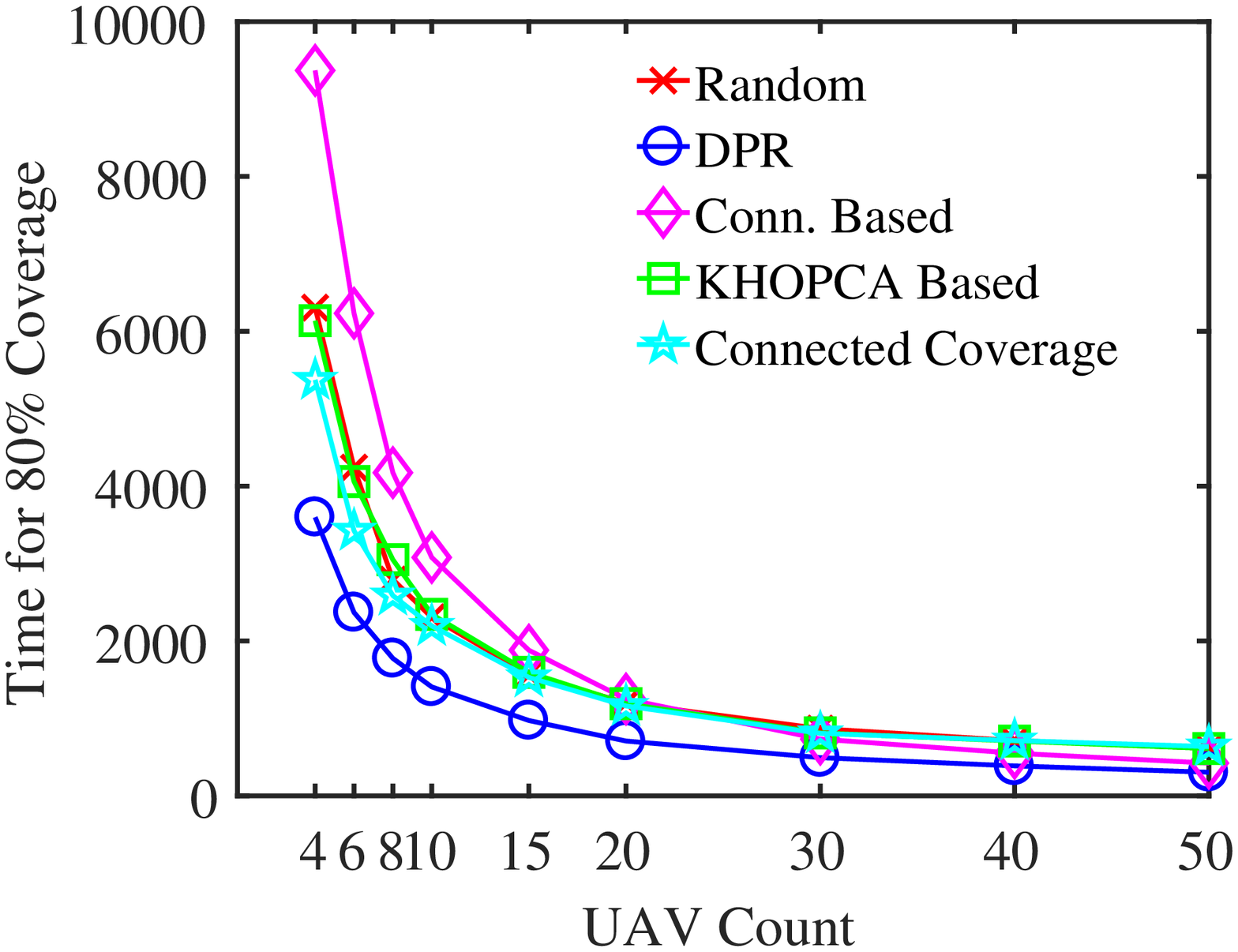}
        \caption{}
    \end{subfigure}%
    ~
    \centering
    \begin{subfigure}{0.5\textwidth}
        \centering
        \includegraphics[scale=0.32]{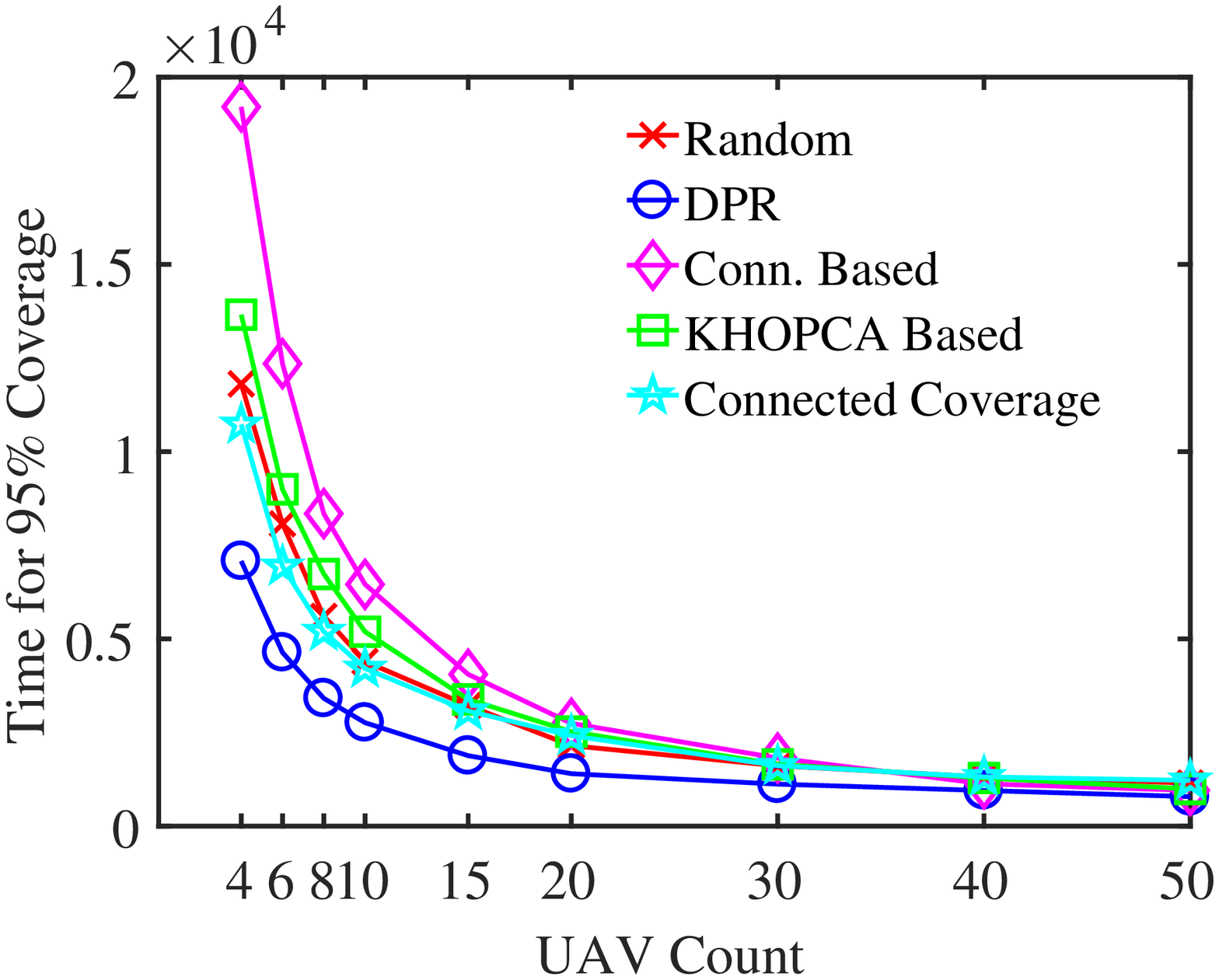}
        \caption{}
    \end{subfigure}
    ~
    \centering
    \begin{subfigure}{0.5\textwidth}
        \centering
        \includegraphics[scale=0.32]{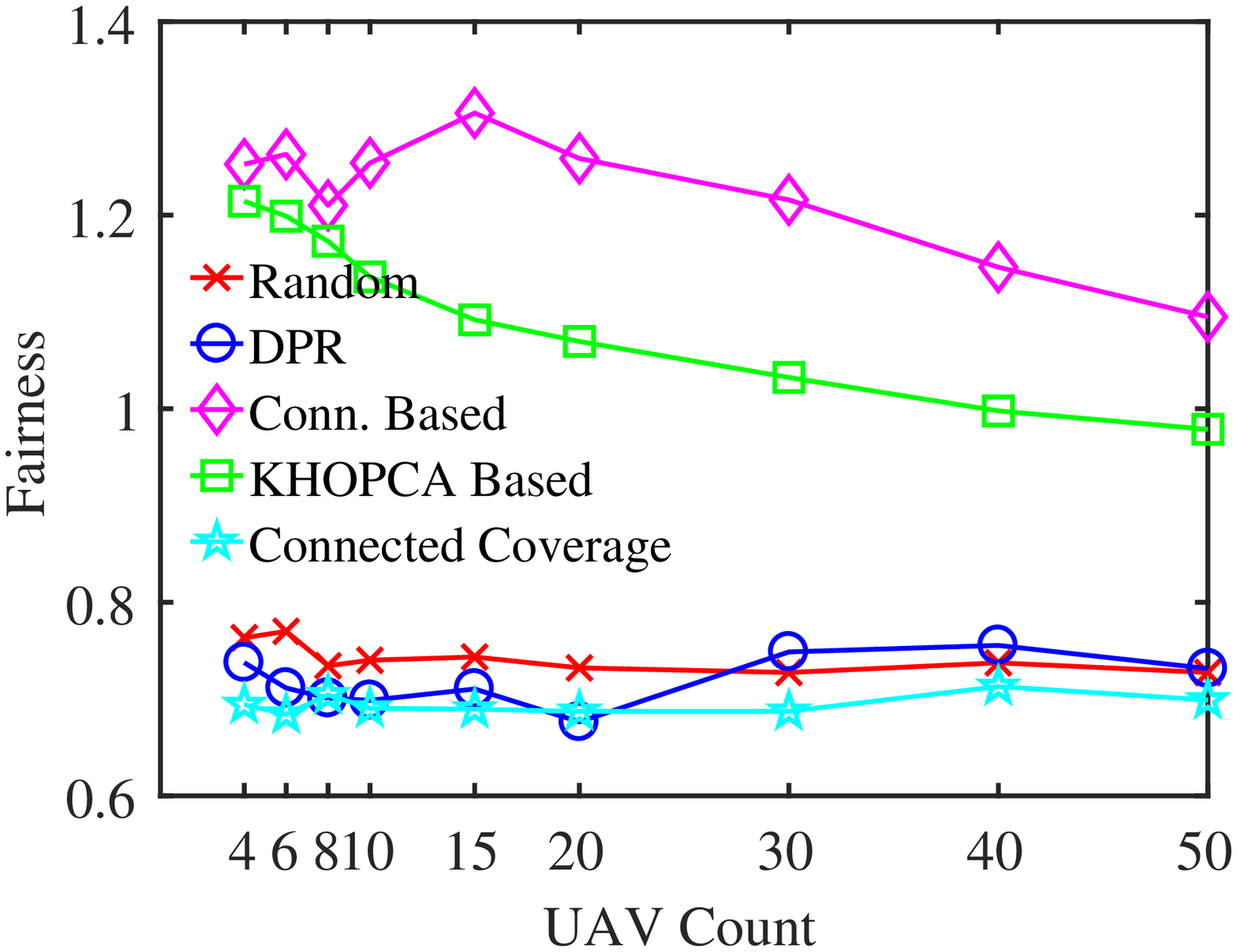}
        \caption{}
    \end{subfigure}
    \caption{Speed of achieving 80\% (a) and 95\% (b) coverage versus number of UAVs. (c) Fairness of coverage versus number of UAVs.}
    \label{cover}
\end{figure}

Figure \ref{cover}a and \ref{cover}b shows the average execution times of the models to cover 80\% and 95\% of the mission field, respectively. According to the results, the time the algorithms need to cover the 80\% of the area is approximately half of the time needed to cover the 95\%, which is an interesting outcome. The DPR has performed better than the other models as expected. Especially in sparser networks, the DPR model is roughly two times faster than the others. The models, that consider connectivity, cover the area more slowly. Moreover, the Connectivity-based and the KHOPCA-based models are even slower than the random-walk model. As the density of the UAVs increases, the ratios between the Connectivity-based model and the others decrease. Even more, the Connectivity-based model has covered the area faster than the other models when the topologies include 30, 40 and 50 drones.

The fairness of coverage policy of the models are given in Figure \ref{cover}c. As it can be seen that the KHOPCA-based and the Connectivity-based models cover the area less balanced than the others. Surprisingly, the random-walk model has given similar results with the DPR and the Connected Coverage models. As the UAV density increases, the Connectivity-based and the KHOPCA-based models scan the area more balanced. The change of the density doesn't affect the fairness of the random-walk and Connected Coverage models dramatically.

\begin{figure}[h!]
    \centering
    \begin{subfigure}{0.5\textwidth}
        \centering
        \includegraphics[scale=0.32]{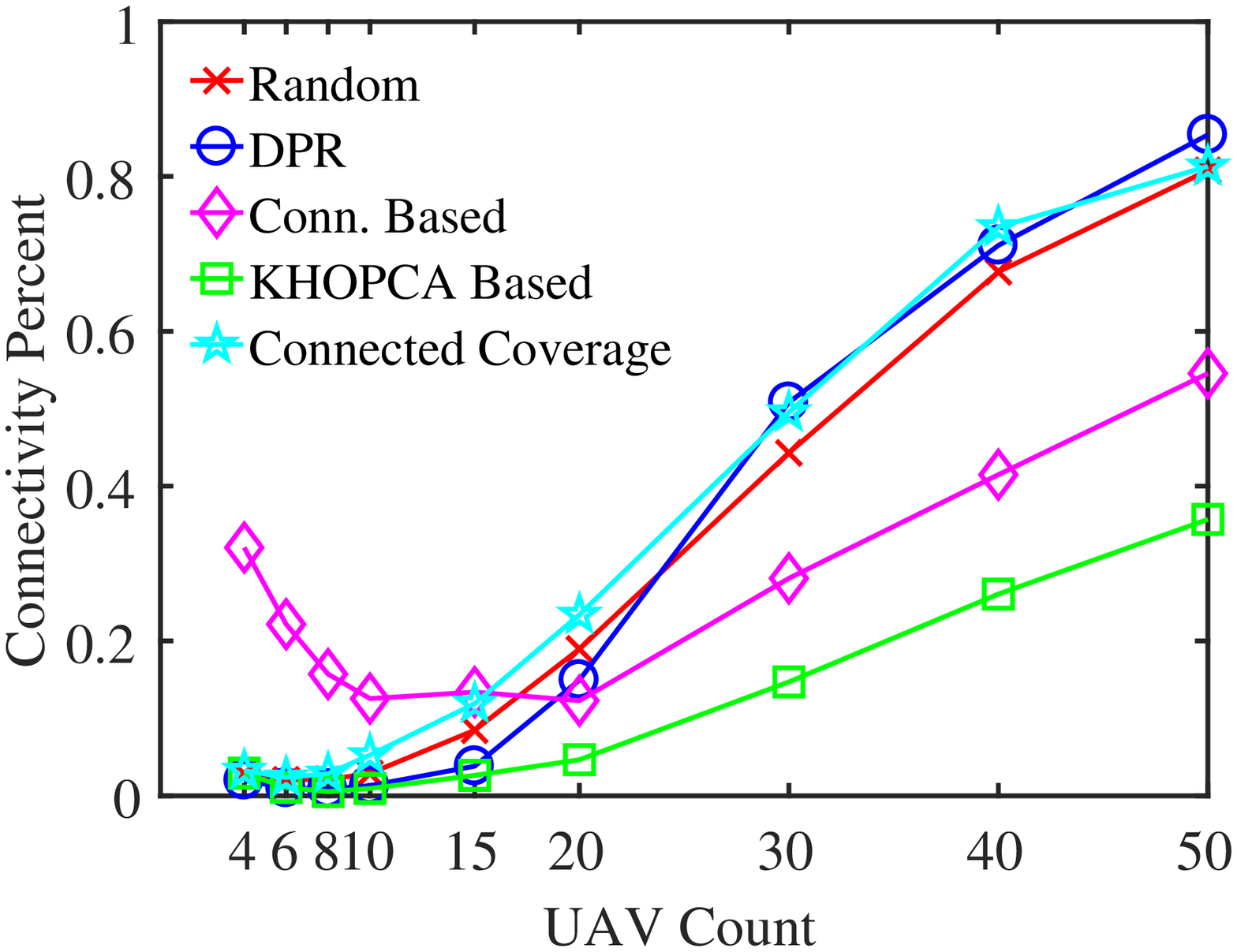}
        \caption{}
    \end{subfigure}%
    ~
    \centering
    \begin{subfigure}{0.5\textwidth}
        \centering
        \includegraphics[scale=0.32]{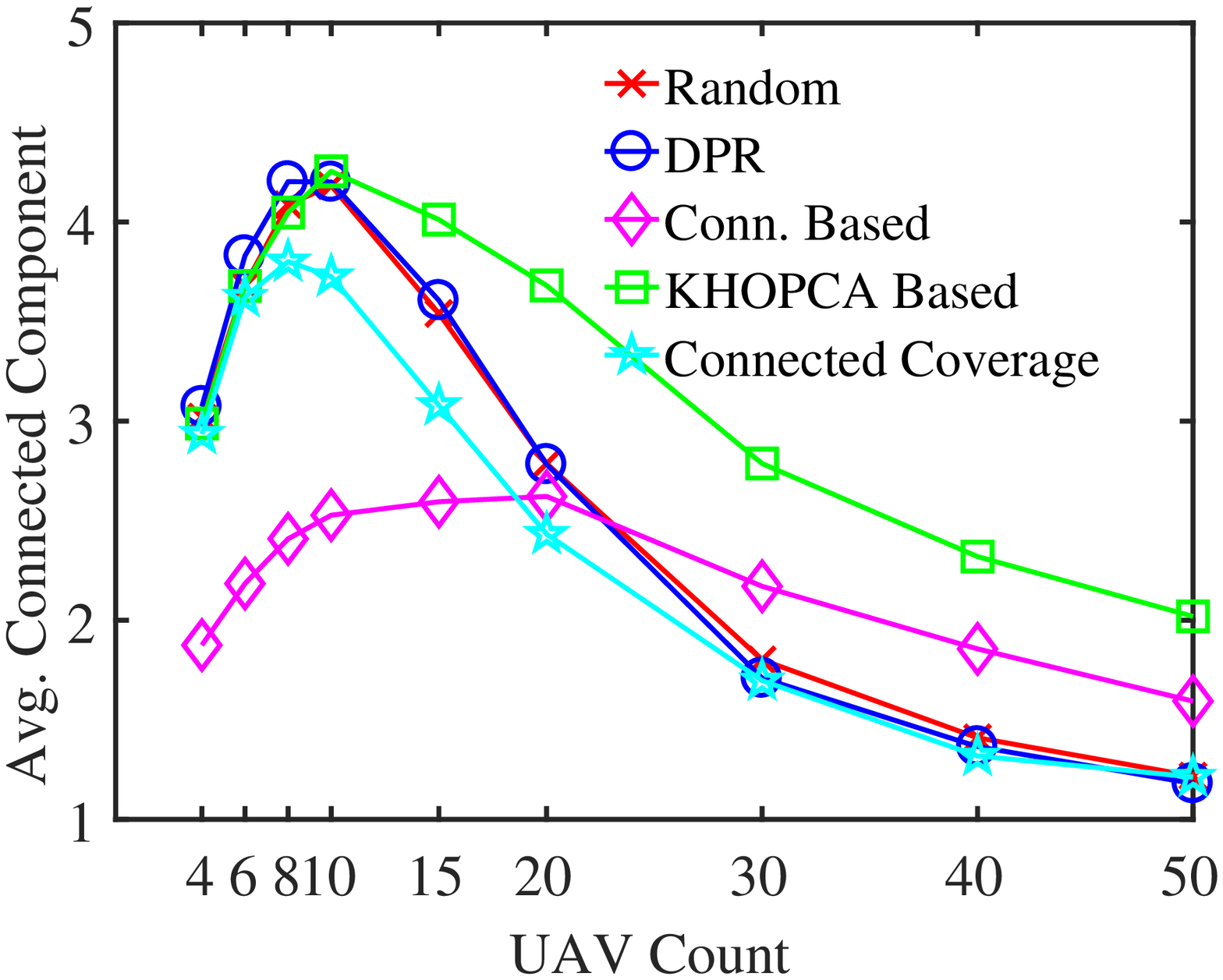}
        \caption{}
    \end{subfigure}
    ~
    \centering
    \begin{subfigure}{0.5\textwidth}
        \centering
        \includegraphics[scale=0.32]{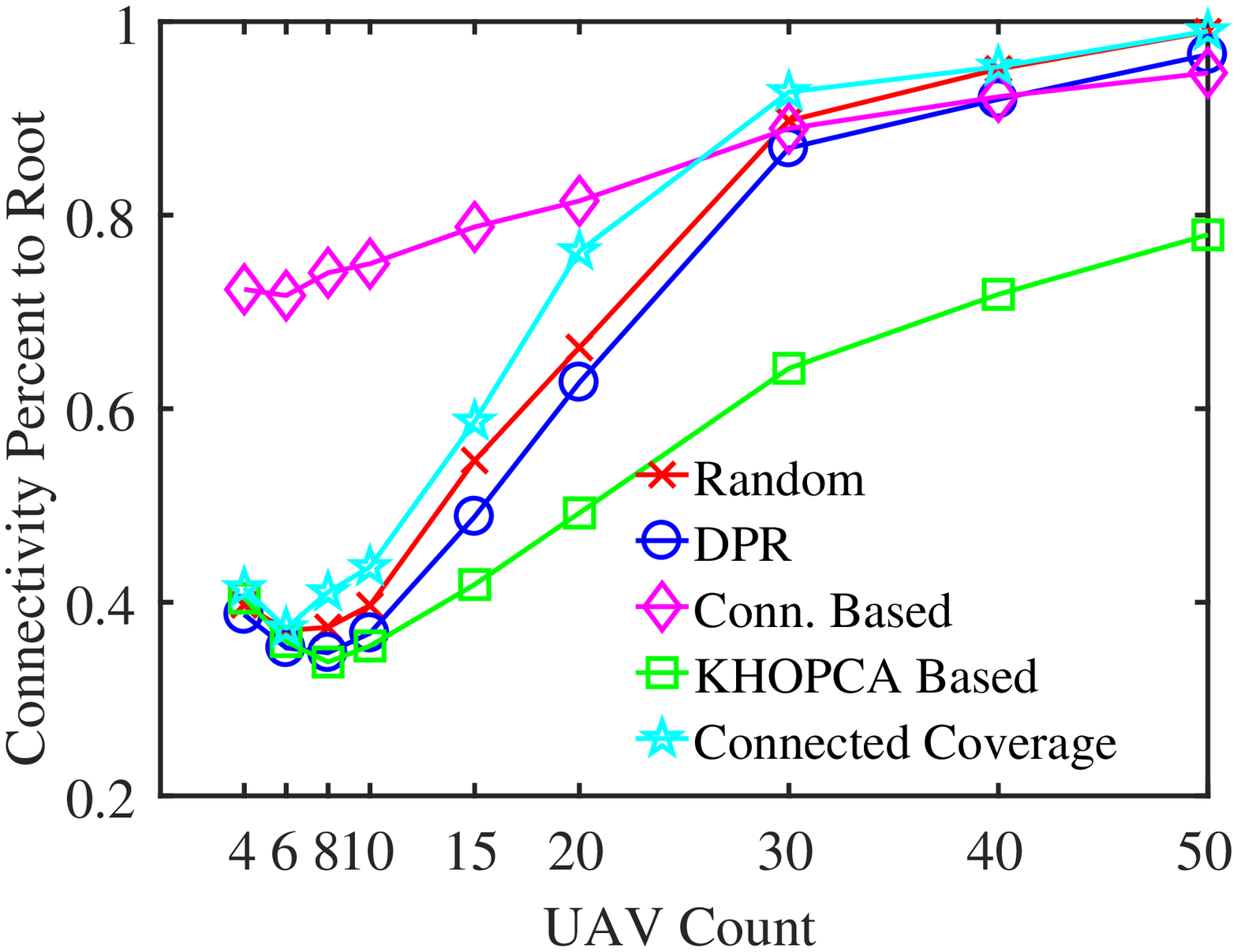}
        \caption{}
    \end{subfigure}
    \caption{(a) Connectivity percentage versus number of UAVs. (b) Average connected component number versus number of UAVs. (c) Connectivity to the root (sink) versus number of UAVs.}
    \label{connectivity}
\end{figure}

Figure \ref{connectivity}a shows that in what percentage of the execution time the UAVs stayed connected within each of the models. Per to the figure, when the number of UAVs is something from 4 to 10, the Connectivity-based model provides much better  connectivity rates than the others. However, as the UAV density increases, its results are getting worse, while the Connected Coverage's, the DPR's and the random-walk's performances are getting significantly better. The average connected component numbers are shown in Figure \ref{connectivity}b. The results are similar to the results presented in the connectivity percentage shown in Figure \ref{connectivity}a, as expected. Because, the more dispersed the network means the less connected the network. The KHOPCA-based model has provided a worse connectivity than the others, as also expected, because it aims to perform clustering over the topology.

Another metric about the connectivity is that the connectivity percentage from any UAV to the root UAV, shown in Figure \ref{connectivity}c. While KHOPCA-based model has provided bad connectivity rates to the root, The Connectivity-based model has provided better, as expected. The DPR, the Connected Coverage and the random-walk models have achieved similar connectivity rates. However, the Connected Coverage model has better connectivity to the root than the DPT and the random-walk models. The Connectivity-based model has provided two times better connectivity to the root than the others in sparse topologies, while they are all similar in dense topologies. 

\begin{figure}[h!]
    \centering
    \begin{subfigure}{0.5\textwidth}
        \centering
        \includegraphics[scale=0.32]{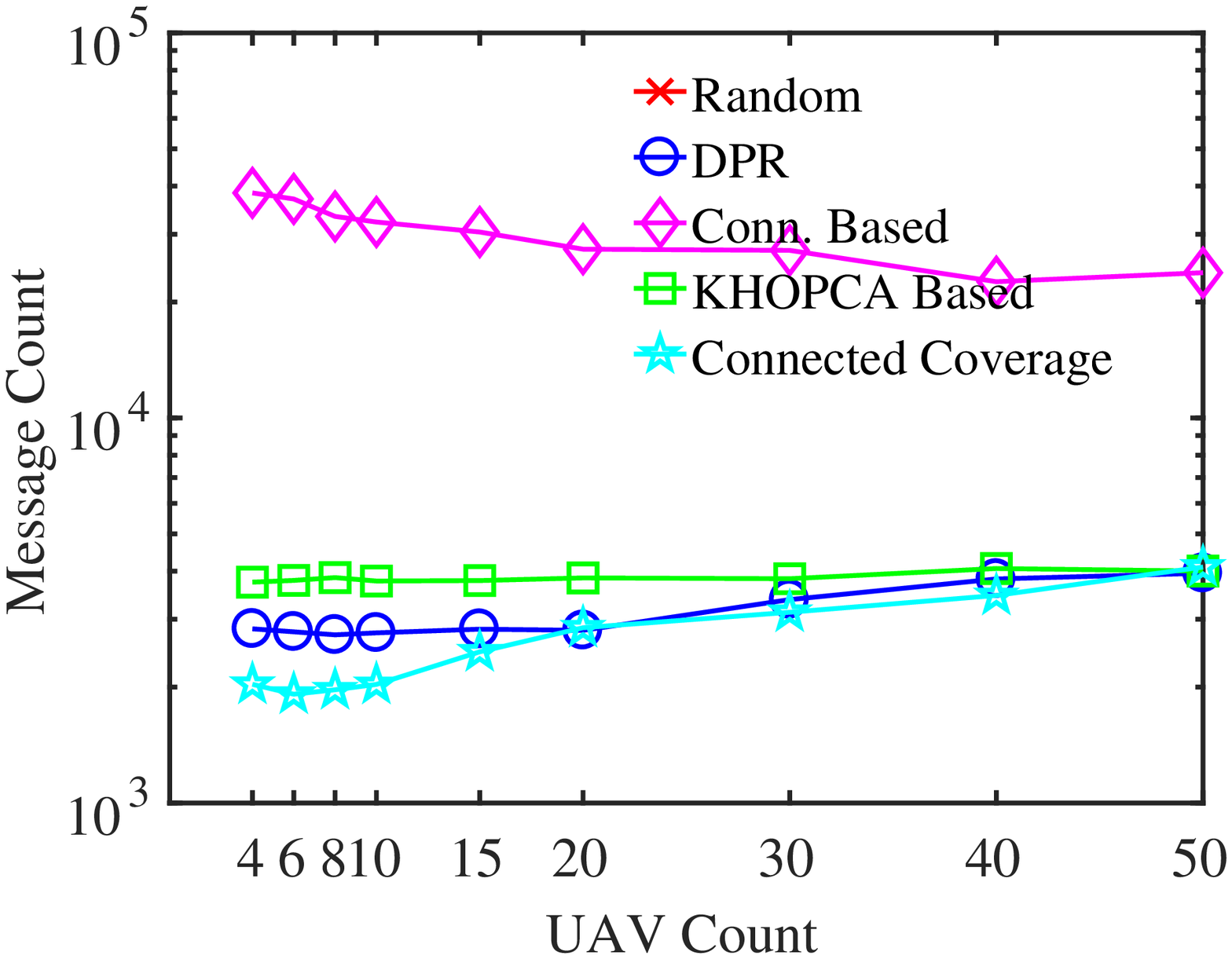}
        \caption{}
    \end{subfigure}%
    ~
    \centering
    \begin{subfigure}{0.5\textwidth}
        \centering
        \includegraphics[scale=0.32]{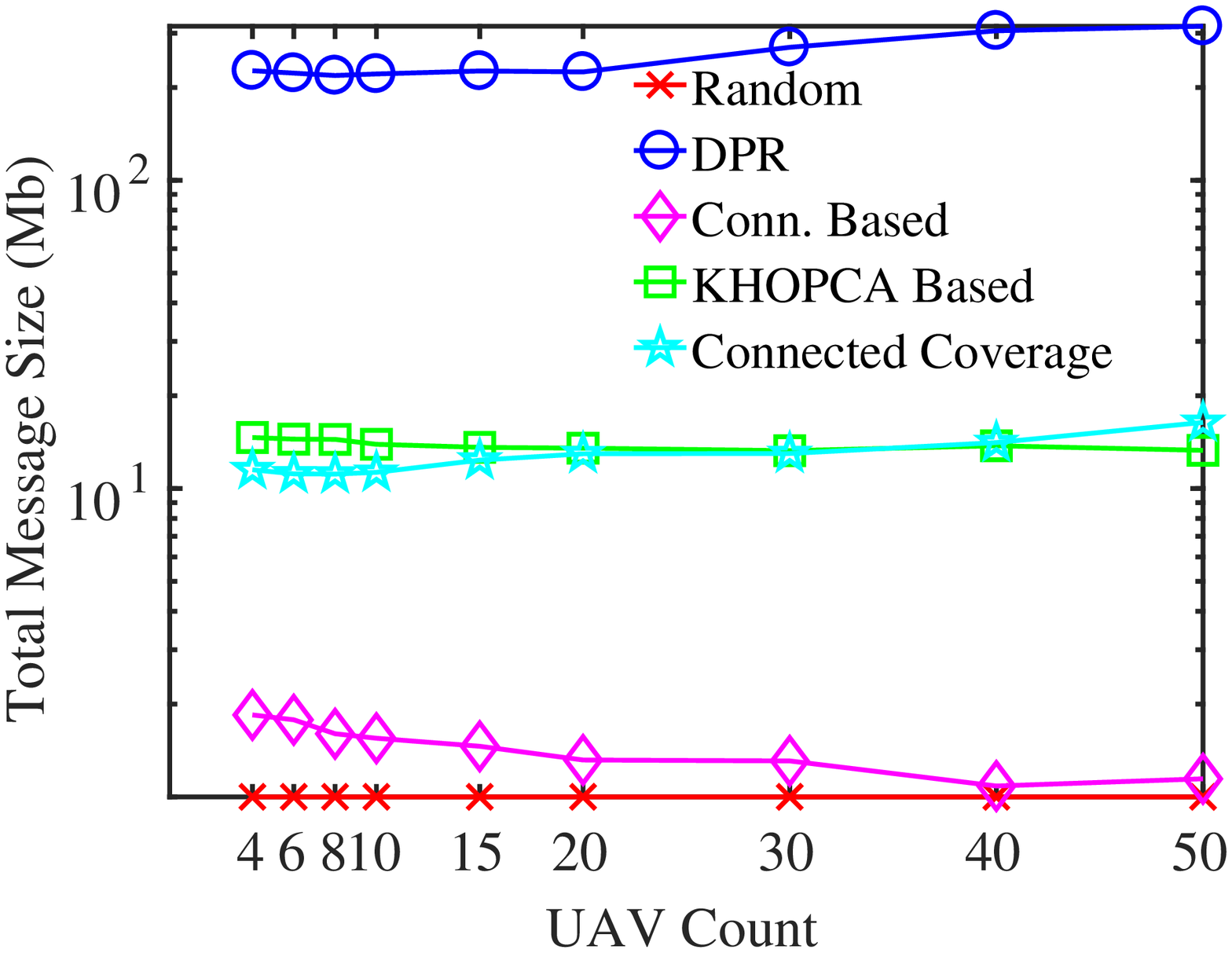}
        \caption{}
    \end{subfigure}
    \caption{(a) Total message counts versus number of UAVs. (b) Cumulative message sizes versus number of UAVs.}
    \label{message}
\end{figure}

The total message counts sent by all UAVs for each model is given in Figure \ref{message}a. The message count is directly related to the decision interval times and the execution times of the models. The shorter interval times or the larger execution times naturally mean more messages to be sent. Since the random-walk model does not need any message to be transmitted, its message number is 0 for all situations. The Connectivity-based model is the one that uses the maximum number of messages due to its short decision interval and long execution time. The Connected Coverage used less messages than the DPR model for 4 to 15 UAVs. For 20 to 50 UAVs, however, they have similar results. The KHOPCA-based model's message count is not affected by the varying UAV numbers, notably. The message counts do not show the precise cost of the communication alone. It is, in fact, required to consider the size of the sent messages, too. Figure \ref{message}b shows the total message sizes sent by the models. Although the Connectivity-based model has sent the highest number of messages, its message sizes are the smallest, because the other models periodically send the maps data sized 100 x 100. Since the DPR cause the UAVs to send maps more frequently than the others, the DPR has the maximum message size. The Connected Coverage and the KHOPCA-based models have comparable message sizes.

\section{Conclusion}

In this study, an extensive performance evaluation of some noteworthy connectivity-aware distributed mobility models tailored for drone networks is presented considering area coverage applications. In accordance, Connectivity-based, KHOPCA-based, Connected Coverage, random-walk and DPR models are implemented. Through simulations, their coverage performances, connectivity qualities, and communication costs are compared over seven designated performance metrics, which are namely the speed of coverage, the fairness of coverage, the connectivity rate, the average connected component number, the connectivity rate to the root, the total message count and the total message volume generated by all UAVs. From a large number of measurements, it is concluded that there is an obvious trade-off between the connectivity rates and the coverage levels. Besides, as it seems, when the models consider the connectivity, their coverage performances are getting lower. According to the results, the DPR model has the best coverage metrics, while the Connectivity-based model pose the best connectivity metrics. The Connected Coverage model uses the least number of messages during its executions, while it generally provides average results over all other performance metrics. In future, it is planned to design and analyze new mobility models, especially to deal with the $k$-connectivity restoration problem in drone networks for arbitrary $k$ values.

%
%
%
\bibliographystyle{abbrv}
\bibliography{ref.bib}

\end{document}